# Dielectric/semiconductor interfacial doping to develop solution processed high performance 1 V ambipolar oxide-transistor and its application as CMOS inverter

Nitesh K. Chourasia[a], Anand Sharma[a], Nila Pal[a], Sajal Biring[b] and Bhola N. Pal[a,b*]

[a]*School of Materials Science and Technology, Indian Institute of Technology (Banaras Hindu University), Varanasi-221005, India,*

[b]*Organic Electronics Research Center and Department of Electronic Engineering, Ming Chi University of Technology, New Taipei City, Taiwan 243*

*\*Corresponding author's e-mail ID: bnpal.mst@iitbhu.ac.in*

**Abstract:**

p-type doping from the dielectric/semiconductor interface of a $SnO_2$ thin film transistor (TFT) has been utilized to develop high carrier mobility balanced ambipolar oxide-transistor. To introduce this interfacial-doping, bottom-gate top-contact TFTs have been fabricated by using two different ion-conducting oxide dielectrics which contain trivalent atoms. These ion-conducting dielectrics are $LiInO_2$ and $LiGaO_2$ respectively, containing mobile $Li^+$ ion. During $SnO_2$ thin film fabrication on top of the ionic dielectric, those trivalent atoms allow p- doping to the interfacial $SnO_2$ layer to introduce the hole conduction in channel of TFT. To realize this interfacial doping phenomena, a reference TFT has been fabricated with $Li_2ZnO_2$ dielectric under the same condition that contains divalent zinc (Zn) atom. Our comparative electrical data indicates that TFTs with $LiInO_2$ and $LiGaO_2$ dielectric are ambipolar in nature whereas, TFT with $Li_2ZnO_2$ dielectric is a unipolar n-channel transistor which reveals the interfacial doping of $SnO_2$. Most interestingly, by using $LiInO_2$ dielectric, we are capable to fabricated 1.0 V balanced ambipolar TFT with a high electron and hole mobility values of 7 cm$^2$ V$^{-1}$ s$^{-1}$and 8 cm$^2$ V$^{-1}$ s$^{-1}$ respectively with an on/off ratio >10$^2$ for both operations which has been utilized for low-voltage CMOS inverter fabrication.

## 1. Introduction

Complementary metal-oxide-semiconductor (CMOS) is a technology for constructing different integrated and analog circuits that are used in microprocessors, microcontrollers, static RAM, image sensors, data converters, and highly integrated transceivers for many types of communication.[1-2] Two important characteristics of CMOS devices are high noise immunity and low static power consumption that makes CMOS as unbeatable technology for modern electronics. This CMOS circuit required complementary and symmetrical pairs of p- and n-channel thin-film transistors (TFTs) for logic functions.[3-5] However, if the deposition conditions of the p- and n- channel TFTs are not compatible, then it is quite difficult to fabricate a highly dense CMOS circuit in a single substrate.[4, 6] In such a scenario, ambipolar TFT can play an important role that shows both electron and hole conduction in a single TFT, based on the gate bias.[6-7] Beside this ambipolar CMOS inverter fabriation, ambipolar charge transport of TFT are widely used for the fabrication light emitting trnasistor,[8-9] flash memery[10-12] and artificialsynaptic emulation.[13] So far, several reports have been shown that individual single-nanostructured based transistors can show good ambipolar behavior,[14-16] but these transistors fail to give a solution for large areas and scalable fabrication. In this contest, thin film-based ambipolar transistors can play a better role and capacity for mass production. To date, most of the reported thin film ambipolar transistor has been fabricated by using organic small molecule and polymer semiconductor.[17-19] However, the main difficulties of these organic/polymer-based TFTs are its low carrier mobility and poor atmospheric stability for electron transport.[6, 11, 19] Relatively, lower-cost metal oxide semiconductors show much high environment stability with higher carrier mobility. Although, it's really hard to find an oxide semiconductor with good hole mobility, which limits the progress of oxide ambipolar TFT fabrication. So far, $SnO_x$ is the only reported oxide semiconductor that is capable of fabricating ambipolar TFT.[20] Although, in most of these cases, it requires a complex device architecture with a control deposition technique. Therefore, for low-cost portable electronics, it's urgent to develop high-performance oxide ambipolar TFT by solution-processed technique with low operating voltage. Those low cost and low voltage ambipolar oxide TFT will be capable to fabricate cost-effective, stable and energy-efficient CMOS inverter for portable electronics.

Lowering the operating voltage of a TFT is a crucial issue for developing low operating voltage CMOS inverter which is required for portable optoelectronics devices. Till now many efforts have been given to developing low voltage TFT by using different high-κ oxide dielectric including $Ta_2O_5$[21], $Y_2O_3$[22], $TiO_2$[23-24], $ZrO_2$[25-26], $HfO_x$[27], $HfLaO_x$[28], $LaAlO_3$[29]. In addition to oxide dielectric, polymer ion-gel and self-assembled monolayer (SAM) have been successfully utilized for low operating voltage organic TFT. However, such kind of ion-gel or SAM dielectric is not compatible with a higher temperature fabrication process, like solution-processed oxide semiconductor. In this situation, ion-conducting oxide dielectric shows the best performance for low operating voltage transparent TFT fabrication.[30-35] Therefore, for high performance with low operating voltage ambipolar TFT fabrication, oxide semiconductor and ionic oxide dielectric is one of the best combination. In addition to that such kind of ambipolar TFT can be environmentally stable due to its intrinsic oxide nature.

Tin oxide is commonly found as an n-type semiconductor. Although, the hole carrier can be introduced to the valance band of $SnO_2$ mainly in two different ways. One of them is to create Sn vacancies which can introduce hole carrier to the valance band of $SnO_2$ which is commonly achieved by sputtering method deposition.[36-38] The other way is by chemical doping with group IIIA elements like indium (In), gallium (Ga) which can occupy a Sn site in the $SnO_2$ lattice.[39-42] Based on those earlier reports, we have fabricated $SnO_2$ TFT by choosing two ion-conducting oxide dielectrics that can introduce p-doping to the dielectric/semiconductor interface to enhanced hole conduction of $SnO_2$. One of them is $LiInO_2$, and the other one is $LiGaO_2$, which can dope In and Ga respectively to an interfacial channel of $SnO_2$. To identify the differences, we have also chosen $Li_2ZnO_2$ as a third ionic dielectric that contains divalent Zn atom, also has no role in introducing hole conduction in $SnO_2$. Comparative electrical characterizations show that all TFTs can show clear n-channel behavior within 2.0 V operating voltage which is due to the common features of oxide semiconductor TFT fabricated on the ion-conducting dielectric.[30-31] However, TFT with $LiInO_2$ and $LiGaO_2$ dielectric show ambipolar behavior whereas a device with $Li_2ZnO_2$ dielectric shows unipolar n-channel behaviour. Moreover, the device with $LilnO_2$ dielectric shows balance carrier transport with high electron and hole mobility within 1.0 V operating voltage, which has been used to fabricate low voltage CMOS inverter.

## 2. Experiment

### 2.1. Synthesis of metal oxide semiconductor and ion-conducting dielectric

The precursor solutions of $LiInO_2$, $LiGaO_2$, and $Li_2ZnO_2$, which were used as an ionic gate dielectric for metal oxide TFT was synthesized by low-cost solution-based synthesis. For the precursor solution of $LiInO_2$ dielectric, a 300 mM solution of lithium acetate has been prepared by dissolved in 2-methoxyethanol. Similarly, 300 mM precursor solution of indium chloride, gallium nitrate, and zinc chloride solution were prepared separately by using 2-methoxyethanol as a solvent. All these solutions except indium chloride were stirred at room temperature for one hour to get the clear solutions of those respective salts. The indium chloride precursor solution took longer time to dissolve which required 24 hours of stirring and has been aged for one day at room temperature before deposition. For the preparation of LiInO$_2$ precursor, solutions of indium chloride and lithium acetate were mixed in a 1:1 ratio and stirred for one more hour at room temperature. Similarly, precursor solutions for LiGaO$_2$ and Li$_2$ZnO$_2$ were prepared by mixing lithium acetate with gallium nitrate and zinc chloride solutions respectively. Finally, a transparent precursor solution of these three ion-conducting dielectrics was used to deposit gate dielectric of our TFTs. A separate 300 mM solution of tin chloride was prepared by dissolving in 2 methoxyethanol followed by continuous stirring for one hour at room temperature that forms a clear solution. This solution was used for the SnO$_2$ semiconducting layer deposition for this TFT.

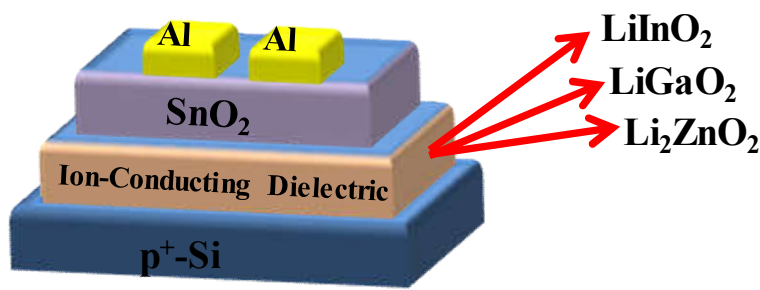

**Figure 1** Schematic view of the device structure

## 2.2. Material Characterizations

### 2.2.1. Thermal Analysis

The thermal behavior LilnO$_2$ was analyzed by thermogravimetric analysis (TGA) and differential thermal analysis (DTA) study. For this experiment, the precursor powder sample was prepared by evaporating the solvent of a mixed solution of lithium acetate and indium chloride. These studies have been performed in a nitrogen (N$_2$) atmosphere at a flow rate of 20 °C /minute. Figure 2 shows the TGA result in combination with DTA in which weight loss ensues in two steps. The first weight loss started from room temperature to 120 °C, corresponding to the residual water, trapped solvent, and moisture which are well supported by DTA peak. This weight loss is ~8 % of the total amount. The second weight loss started from 470 °C and ended with 550°C and the sharp intense DTA exothermic peak is observed in this temperature range which indicates crystallization of sol-gel $LiInO_2$ powder. There is a negligible weight loss between the temperatures range of 550°C to 800°C. This nature of DTA suggests the crystallization process of $LiInO_2$ is occurring ~550 °C at the highest rate. Thermal behavior of $LiGaO_2$ is shown in Figure S1 which indicated the crystallization of this dielectrics is 550 °C and the crystallization temperature of dielectric $Li_2ZnO_2$ is also 500 °C which has been reported in our earlier paper.[33]

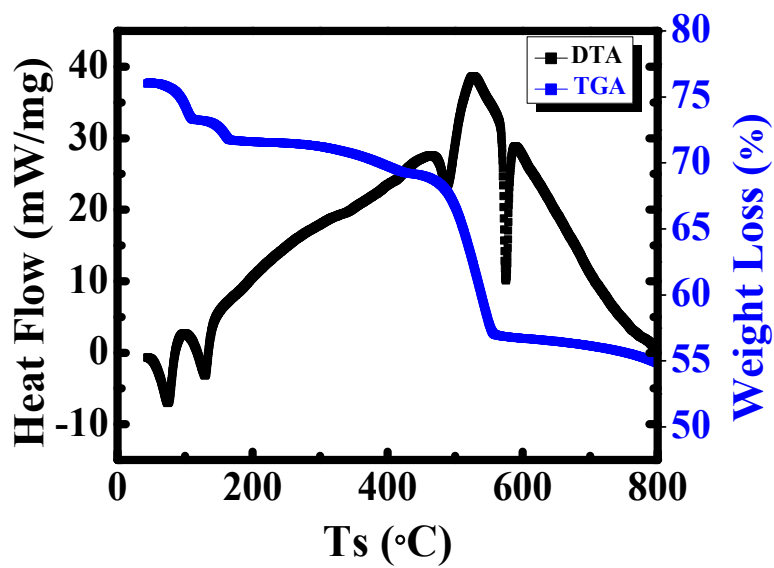

**Figure 2** Thermal gravimetric analysis (TGA) and differential thermo-gravimetric analysis (DTA) curves of precursor powder $LiInO_2$

*2.2.2. Structural Properties*

Grazing Incidence X-ray Diffraction (GIXRD) measurement was used to check the structural property of $LiInO_2$ thin film. For this measurement, the thin film sample was prepared on a glass substrate and annealed at 550 °C for 1 hr. Fig. 3 a) shows the GIXRD data of $LiInO_2$ thin film annealed at 550 °C. The diffraction peaks that originated from reflection planes of (103), (211) and (204) at peak 2θ angles near about 35.51, 48.14 and 58.18 respectively, recognized the tetragonal crystal phase of $LiInO_2$ which has been verified by JCPDS data (No. 76-0427). These sharp diffraction peaks suggest that 550 °C annealing temperature is enough to achieve the crystalline phase of $LiInO_2$ dielectric thin film. The GIXRD data for $LiGaO_2$ and $Li_2ZnO_2$ thin films are shown in Figures S2 and S3 respectively. Figure S2 shows that the $LiGaO_2$ is crystalline in nature and preferred crystalline orientation from reflection planes of (002) and (040) at peak 2θ angles near about 35.8 and 57.8 respectively.[43] Similarly from Figure S3, clear reflection from (120) and (112) planes are observed for $Li_2ZnO_2$ that has been reported in our earlier work. [33]

Fig. 3 b) represents Grazing Incidence X-ray Diffraction (GIXRD) analysis of $SnO_2$ thin film prepared on a glass substrate annealed at 500 °C for 30 minutes. The diffraction peak of $SnO_2$ were assigned as $26.56^0$, $33.92^0$, $38.08^0$ and $51.72^0$ originated from reflection planes (110), (101), (200) and (211) respectively, suggest the tetragonal phase of $SnO_2$.[44]

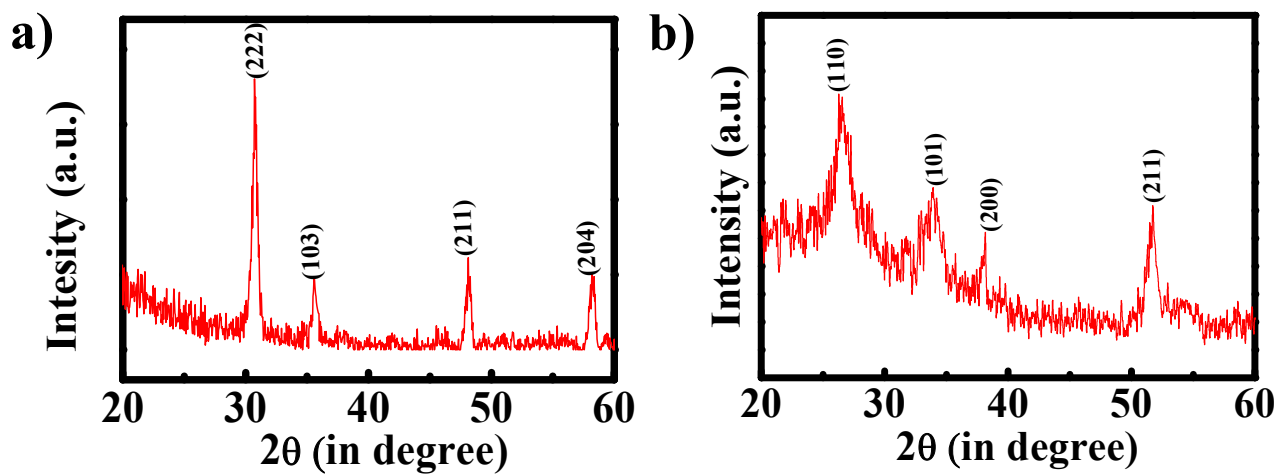

**Figure 3 a)** GIXRD analysis of $LiInO_2$ thin film at $550^0C$ annealing temperature and **b)** GIXRD analysis of $SnO_2$ thin film at $500^0C$ annealing temperatures

*2.2.3. Surface Morphology*

For TFT, the dielectric/semiconductor interface plays a crucial role in device performance. Therefore, atomic force microscopy (AFM) was used to study the surface morphology of dielectric thin film ($p^+$-Si/ $LiInO_2$) annealed at 550°C as shown in the Figure 4). From AFM exploration, the root means square (rms) value of $LiInO_2$ thin film is extracted approximately 5 nm which is acceptable for solution-processed TFT. The 3-D image of $LiInO_2$ thin-film AFM also suggests that the film is dense and void-free. Similarly, AFM studies of $LiGaO_2$ and $Li_2ZnO_2$ thin films have been performed which are shown in Figure S4 and S5 respectively. These two pictures indicated the rms roughness for these two films are ~3 and 6 nm respectively.

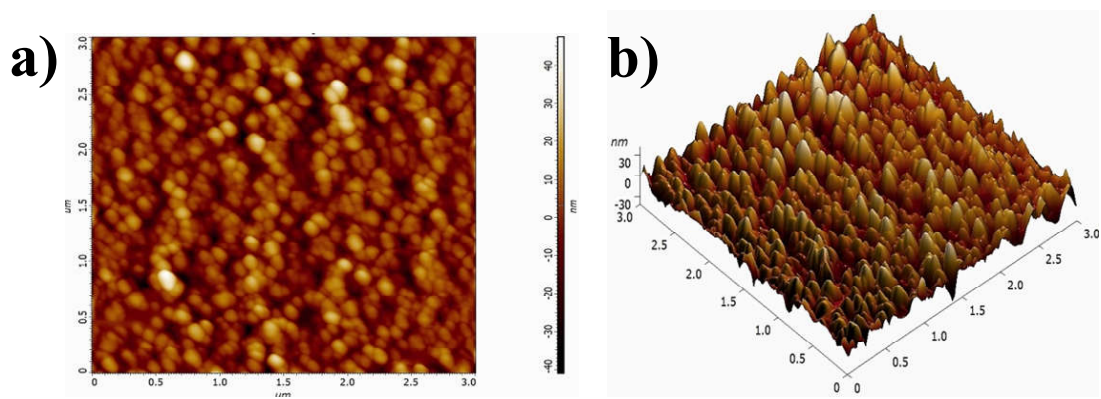

**Figure 4** Surface morphologies (scan surface area 3 x 3 μm) of the solution-processed $LiInO_2$ dielectric thin films for $LiInO_2/p^+$-Si surface annealed at 550 °C **a)** 2-D topography **b)** 3-D topography

*2.2.3. Optical Properties of LiInO₂ Thin Films*

The optical transmittance spectra of sol-gel derived $LiInO_2$ thin film was recorded in the wavelength range 300–900 nm. Figure 5 a) showed the spectral transmittance spectra of $LiInO_2$ film coated on a quartz substrate that was annealed at 550°C for one hour. It was noticed that the dielectric sample has low transmittance in the ultra-violet region (300–400 nm) but high average transmittance (≈ 84%) in the visible region (400–850 nm). Higher transmittance in the visible region specifies the dielectric film is very smooth with low defect density and voids, which is very beneficial for the fabrication of high-performance TFT with low leakage current. For that reason, $LiInO_2$ thin film can be a suitable candidate to use as a gate dielectric for high-performance TFT. The optical band gap energy of the $LiInO_2$ thin film sample was calculated by extrapolating linear region of the plot $(αhv)^2$ versus hv (Figure 5 b), where hv is the incident photon's energy and $α$ is the optical absorption coefficient. The extracted value of the energy band gap of $LiInO_2$ is 3.6 eV, which is the same as previously reported.[45] The optical transmittance spectra and Tauc's plot of $LiGaO_2$ thin films are given in Figure S6 which looks very similar to nature with $LiInO_2$. However, the extracted band gap of $LiGaO_2$ is 5.5 eV which is much higher than $LiInO_2$ thin film. However, the optical band gap of $Li_2ZnO_2$ is much lower (3.3 eV) which has been reported in our earlier work.[33]

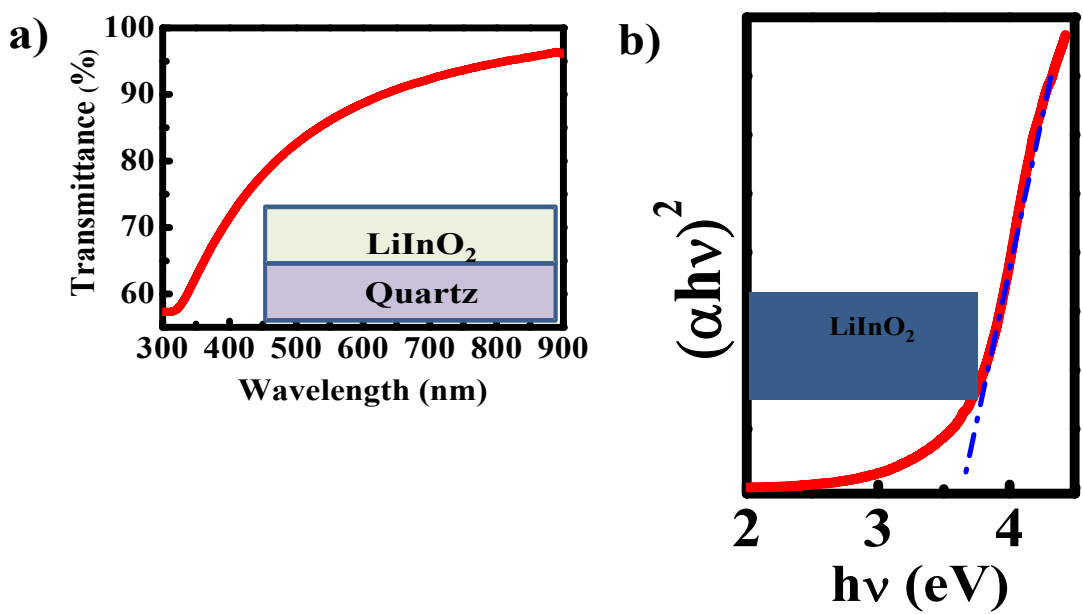

**Figure 5 a)** Optical transmittance spectra of the solution-processed $LiInO_2$ dielectric thin film annealed at 550°C $LiInO_2$/quartz(inset) **b)** The Tauc's plot corresponding to $LiInO_2$ dielectric thin film

*2.2.4. Device fabrication*

Three different types of TFTs have been fabricated by using $LiInO_2$, $LiGaO_2$, and $Li_2ZnO_2$ dielectric thin film, respectively, with a bottom-gate top-electrode geometry which we called as TFT1, TFT2, and TFT3 respectively. All these devices have been fabricated on top of heavily doped p-type Si ($p^+$-Si) substrates of dimension 15 mm x 15 mm. In the beginning, all these substrates were passes through routine cleaning in four different solutions as previously reported.[32] After routine cleaning, all the substrates were dried by passing dry air and immediately exposed to oxygen plasma for 5 min to remove unwanted organic residue (hydrocarbons) from the Si substrate and make the substrate hydrophilic. Such a hydrophilic surface offers smooth and pinhole-free thin film formation during spin coating which supports diminishing the trap state on the dielectric surface. Before spin coating, all precursor solutions were filtered through a syringe filter (PVDF -0.45μm) due to which film quality was improved. Afterward, the solution of the dielectric precursor of $LiInO_2$ was spin-coated at 4000 rpm for 50 seconds on the top of Si substrates under ambient atmospheric conditions. To remove the precursor solvent, the spin-coated film was kept on a hot plate at 80 $^o$C for two minutes followed by the annealing process (350 $^o$C) in a muffle furnace for half an hour. This process was repeated two more times. Finally, the dielectric thin film coated on Si substrate was annealed at 550 $^o$C in a furnace for one hour to obtain the polycrystalline phase of $LiInO_2$ under ambient atmospheric condition. Similarly, for TFT2 and TFT3, $LiGaO_2$ and $Li_2ZnO_2$ thin films are deposited in a similar process by three successive coatings followed by annealing process at 550 $^0$C and 500 $^0$C for one hour and half an hour , respectively. Solution-processed tin oxide ($SnO_2$) that was used as a metal oxide semiconductor of TFTs, was coated on top of the ionic gate dielectric. In this deposition process, a 300 mM precursor solution of $SnO_2$ was spin-coated onto three different ion-conducting dielectric films for three different types of TFTs fabrication which were kept on a preheated hot plate at 120 $^o$C for 2 minutes for drying. After that, dried thin films were transferred to a furnace for annealing the film at 500 °C for half an hour to obtain a polycrystalline film of $SnO_2$ on the dielectric surface. Finally, aluminum metal has been deposited on the top of $SnO_2$ thin film by shadow mask process in a thermal evaporator that works as a source and drain electrodes of the devices. The width to length ratio of the TFT was 118 (W/L=23.6 mm/0.2 mm).

*2.2.4. Dielectric and Electrical characterizations*

To understand the electrical properties of the deposited dielectric thin films, current-voltage (I - V) measurements have been carried out using a metal-insulator-metal device architecture ($p_+$-Si/dielectric/Al). The leakage current density of $550^0C$ annealed $LiInO_2$ dielectric thin film at 2 V is only 2.4 x $10^{-8}$ Amp/ $cm^2$ which is very low as compared to other previously reported dielectric (Figure 6a).[32-33] This low leakage current density is the signature of the highly-dense dielectric thin film with very less defect density. Apart from this, the breakdown voltage of the device is ~ 16 V, which is around sixteen-time higher than the normal operating voltage of the device ($\leq 1$ V). Thus, the above observations are fruitful evidence to use $LiInO_2$ as a gate dielectric for ambipolar TFTs.

The dielectric behavior of $LiInO_2$ was examined in detail with the same device structure ($p^+$-Si/$LiInO_2$/Al) by measuring frequency-dependent capacitance (C - $f$) within the range of 20 Hz to 1 MHz as shown in Figure 6 b). The capacitance of the $LiInO_2$ film decreases with frequency particularly above $10^3$ Hz because of its strong depends on ionic polarization due to the movement of $Li^+$ which is a relatively slow process. The measured capacitance per unit area of fabricated $LiInO_2$ thin film is 478 $nF/cm^2$ which is significantly higher than the thermally grown $SiO_2$ with similar thickness. This higher capacitance per unit area value of $LiInO_2$ thin film indicates its suitability as a gate dielectric for low operating voltage TFT fabrication. Similarly, frequency-dependent capacitance characterization has been performed for $LiGaO_2$ and $Li_2ZnO_2$ thin films, those are shown in Figure S7 and S8, respectively. The measured capacitance per unit area for LiGaO$_2$ thin film is 350 nF/cm$^2$, and Li$_2$ZnO$_2$ thin film is 312 nF/cm$^2$.

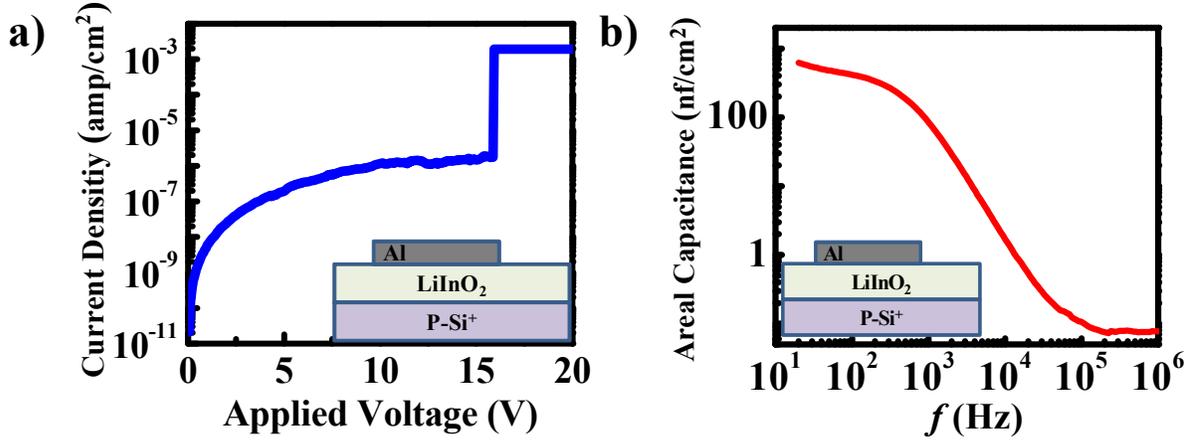

**Figure 6 a)** Leakage current density vs applied voltage of LiInO$_2$ thin film annealed at 550 °C with p$_+$-Si/LiInO$_2$/Al device structure **b)** capacitance versus frequency curves of solution-processed ionic dielectric LiInO$_2$ thin film annealed at 550 °C

To identify the device performance of three ion-conducting oxides (LiInO$_2$, LiGaO$_2$, and Li$_2$ZnO$_2$) as a gate dielectric, three different TFTs were fabricated on highly doped Si (p$^+$-Si) substrate using polycrystalline SnO$_2$ as a channel semiconductor (Figure 1). As mention earlier, we named these three TFTs with LiInO$_2$, LiGaO$_2$, and Li$_2$ZnO$_2$ dielectrics as TFT1, TFT2 and TFT3 respectively. Fig. (7) Shows the output and transfer characteristics of all three TFTs under low voltage operation. The applied gate voltage was swept from -0.5V to 2V for n-channel operation with drain voltage variation from 0 V to 1 V. On the other hand, for p-channel operation, gate voltage was varied from 0.5 V to -2 V with drain voltage variation from 0V to -1V. Figure 7a) and 8a) show the output characteristics of TFT1 under n-channel and p-channel operation, respectively which shows drain current (I$_D$) saturate under <1.0 V operation which is advantageous for low power electronics. Additionally, these data indicate that the nature of drain current amplification under n- and p-channel operation is very much similar with same threshold voltage that results in almost the same saturation drain current at $|V_g|=2$ V for both types of operation which is the signature of the balanced ambipolar transistor. Fig.7 (d) and 8(c) shows the transfer characteristics for n- and p- channel operation of TFT1 and the drain current increases with either positive (for n-channel) or negative gate bias (for p-channel) under $|V_d|=1$ V which implies that this transistor "turns on" under both types of gate bias. Similar characterizations have been done for TFT2 and TFT3. Figure 7b) and Figure 8b) shows the output characteristics of TFT2 for n-channel and p-channel transport respectively. Similarly, Figure 7e) and Figure 8d) shows the transfer characteristics for n-channel and p-channel transport respectively. These data indicated that TFT2 shows dominating n-channel transport and weak p-channel transport. On the other hand, TFT3 shows pure n-channel transport which has been identified from their output and transfer characteristics that are shown in Figure 7c) and Figure 7f), respectively. Instead of having their different charge transport nature, output characteristics of all these TFTs show good current saturation bellow 2 V operating voltage, indicates common noble features of the ion-conducting oxide gate dielectric.

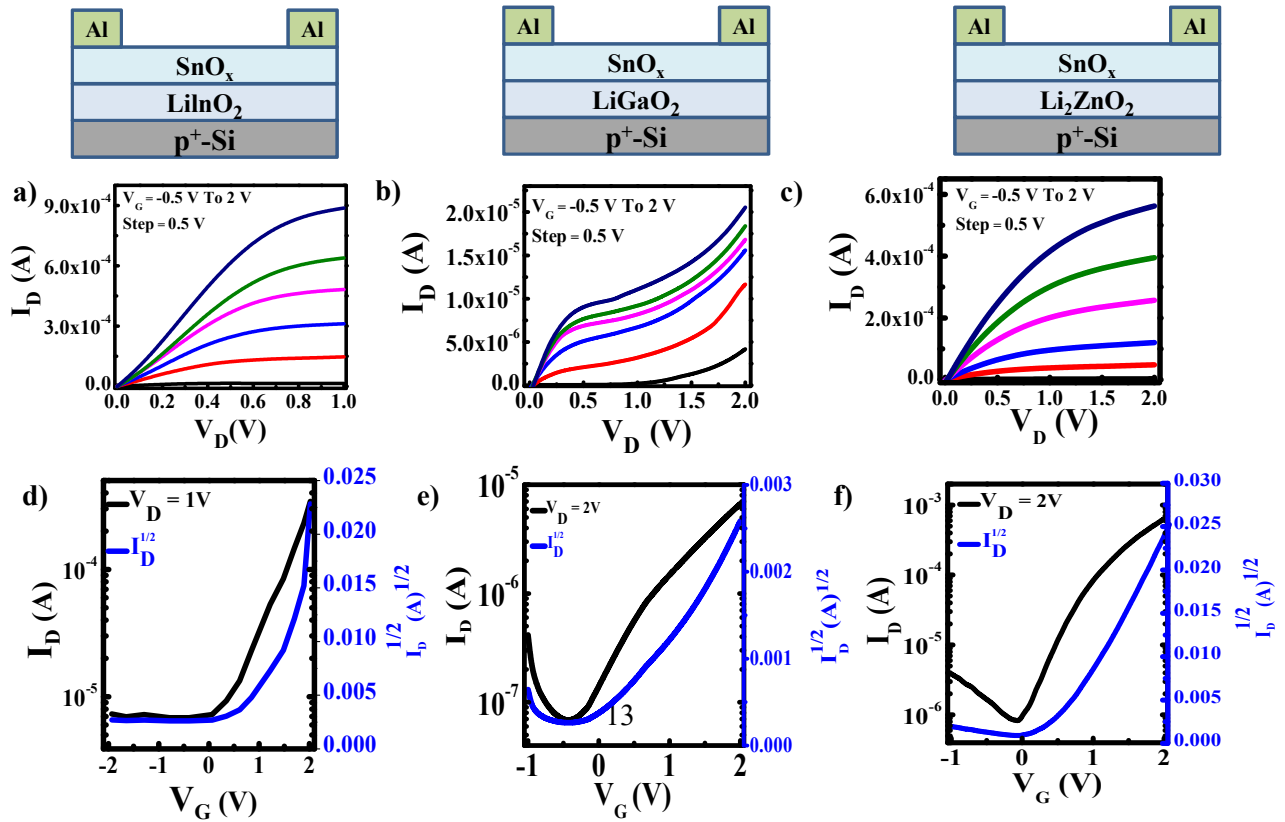

**Figure 7 a)** Output and **d)** transfer characteristics of the SnO$_2$ TFT with LiInO$_2$ dielectric annealed at 550$^0$C, **b)** Output and **e)** transfer characteristics of the SnO$_2$ TFT with LiGaO$_2$ dielectric annealed at 550$^0$C and **c)** Output and **f)** transfer characteristics of the SnO$_2$ TFT with Li$_2$ZnO$_2$ dielectric annealed at 500$^0$C with device architecture Al/SnO$_2$/LiInO$_2$/p$^+$-Si, Al/SnO$_2$/LiGaO$_2$/p$^+$-Si and Al/SnO$_2$/ Li$_2$ZnO$_2$/p$^+$-Si respectively, under n-channel operation

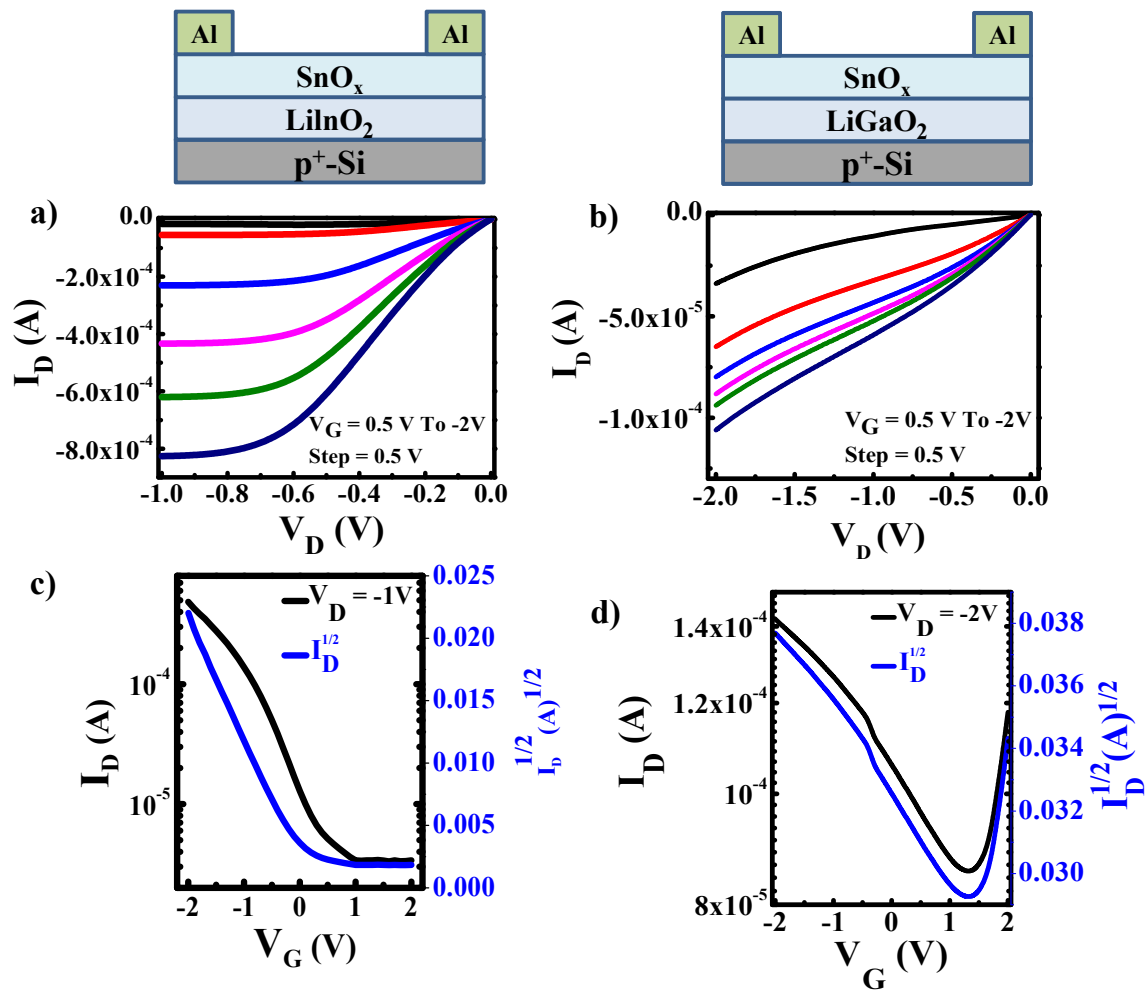

**Figure 8: a)** Output and **c)** transfer characteristics of the SnO$_2$ TFT with LiInO$_2$ dielectric annealed at 550$^0$C, **b)** Output and **d)** transfer characteristics of the SnO$_2$ TFT with LiGaO$_2$

dielectric annealed at 550°C with device architecture Al/SnO$_2$/LiInO$_2$/p$^+$-Si, and Al/SnO$_2$/LiGaO$_2$/p$^+$-Si under p-channel operation

## 4. Results and discussion

The effective mobility of the carrier (μ) and sub-threshold voltages (SS) of this TFT is calculated from the following equations;

$$I_D = \mu C \frac{W}{2L}(V_G - V_T)^2 \quad \text{.....................} \quad (1)$$

$$SS = \left[\frac{d(\log I_D)}{dV_G}\right]^{-1} \quad \text{.....................} \quad (2)$$

Where $I_D$, C, $V_G$, $V_T$ are saturation drain current, capacitance per unit area, gate voltage and threshold voltage. Since the TFT operation was performed in direct voltage, the capacitance at a lower frequency (50 Hz) is taken into account for the calculation of mobility to avoid overestimation. The threshold voltage of the device can be calculated by fitting a straight line on $I_D^{1/2}$ vs. $V_G$ plot of transfer characteristics. The extracted carrier mobility and threshold voltage for n-channel and p-channel operations of TFT1 are 7 cm$^2$ V$^{-1}$ s$^{-1}$, 0.2V and 8 cm$^2$ V$^{-1}$ s$^{-1}$, 0.3V respectively. To the best of our knowledge, as an ambipolar transistor, these are the highest achieved mobilities among all reported organic/polymer or oxide devices.

Moreover, those above mobility suggests that there is a balance injection of both types of charge carriers (i.e., electron and hole) which is very necessary for CMOS electronics, are not commonly found in earlier organic/polymer or inorganic oxide-based ambipolar TFT. In addition to carrier mobility, two other parameters determine TFT device quality. One of them is the current on/off ratio and the other one is the device's sub-threshold swing (SS). The on/off ratio of the device with n-channel and p-channel operation is 60 and 1.5 x 10$^2$, respectively. On the other hand, the subthreshold swing of the device is 1.31 V dec$^{-1}$ (for n-channel) and 0.97 V dec$^{-1}$ (for p-channel), respectively. Again as an ambipolar TFT, these on/off ratio are quite high and SS values are significantly low, which are required CMOS inverter circuit. In addition to TFT1, TFT2 also shows a very good n-channel behavior with electron mobility, on/off ratio and ss values 0.35 cm$^2$ V$^{-1}$ s$^{-1}$, 10$^2$ and 1.46 V/decade, respectively. However, this device show relatively poor p-channel behavior with a hole mobility of 0.59, on/off ratio of 2 and ss value of 1.12 V/decade. In contrast, TFT3 shows only n-channel behavior with electron mobility of 16 cm$^2$ V$^{-1}$ s$^{-1}$, on/off ratio of 10$^3$ and SS value of 0.41 V/decade which are very good as n-channel transistor but doesn't show any p-type transport in the channel. The device parameters of all these TFTs are summarized in table-1. The comparative study of these three different TFTs indicates that the device with LiInO$_2$ dielectric shows the best ambipolar field-effect transistor behavior with very high mobility at the low operating voltage with good on/off ratio and low subthreshold swing. This outstanding performance of TFT1 clearly indicates that this combined dielectric/semiconductor device architecture great potential to give a big boost of ambipolar TFT research.

To realised this comparative electrical behavior, we proposed a dielectric/semiconductor interfacial doping phenomena which has been descrived in Figure 9. As per the earlier report, group IIIA elements work as an acceptor of SnO$_2$ semiconductor.[46] Out of these different group IIIA elements, In and Ga have been theoretically predicted to introduce deep acceptor with an activation energy of 580 and 760 meV (Figure 9 a) and 9 b)). [47-48] Besides, experimental reports also show that the initial addition of In or Ga doping increases the resistance of the SnO$_2$ semiconductor by two orders of magnitude. During this period, conduction due to electron steeply decreases and hole conduction introduces which reaches the peak value for the highest level of doping (~10$^{17}$/cm$^2$).[47] However, after crossing the maximum level of doping concentration, resistivity rapidly decreases and hole conduction disappears which is due to the formation of In$_2$O$_3$ or Ga$_2$O$_3$ secondary phases in the SnO$_2$ thin film that works as a pure n-type semiconductor.[47] In our present work, during the annealing process of SnO$_2$ semiconductor, In or Ga has been thermally diffused to the SnO$_2$ through the SnO$_2$/ion-conducting dielectric interfaces which introduced deep acceptor to the interfacial SnO$_2$ (Figure 9 c).[49] Again, as it is known that the conducting channel of TFT is formed in the dielectric/semiconductor interfacial channel of the device.[6, 50] Therefore, this deep acceptor introduce hole conduction to the channel of TFT1 and TFT2. As it mentions earlier, acceptor activation energy due to In doping is relatively lower than Ga,[47] (Figure 9a) & b)) therefore, compared to TFT2, TFT1 shows better hole conduction and shows a more balanced electron and hole transport. However, this acceptor doping can not be introduced by group IIB Zn atom.[33]. Thus, TFT3 doesn't show any hole conduction.

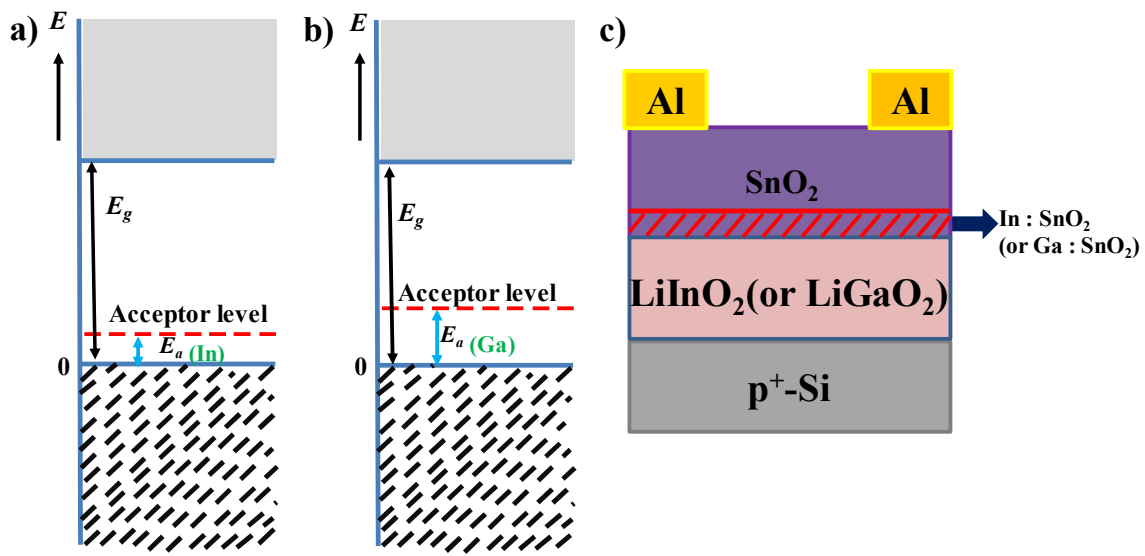

**Figure 9)** Schematic of deep acceptor level with tin oxide (SnO$_2$) for **a)** In , **b)** Ga and **c)** schematic presentation of dielectric/semiconductor interfacial doping

Table-1. The summary of different device parameters of three SnO$_2$ TFTs fabricated with three different ion-conducting oxide dielectrics

| Device no. | Dielectric | Dielectric Annealed temperature | C (nF/cm$^2$) at 50 Hz | V$_{Th}$ (V) | ON/OFF | Subthreshold swing (SS) (V/decade) | Mobility ($\mu$) (cm$^2$ V$^{-1}$ sec$^{-1}$) |
|---|---|---|---|---|---|---|---|
| 1. | LiInO$_2$ (n-operation) | 550 °C | 478 | 0.2 | 60 | 1.31 | 7 |
| 2. | LiInO$_2$ (p-operation) | 550 °C | 478 | 0.3 | 1.5 x 10$^2$ | 0.97 | 8 |
| 3. | LiGaO$_2$ (n-operation) | 550 °C | 350 | -0.4 | 10$^2$ | 1.46 | 0.35 |
| 4. | LiGaO$_2$ (p-operation) | 550 °C | 350 | 1.1 | 2 | 1.12 | 0.59 |
| 5. | Li$_2$ZnO$_2$ (n-operation) | 500 °C | 312 | 0.3 | 10$^3$ | 0.41 | 16 |

To verify the applicability of this SnO$_2$ based TFTs for CMOS circuit, an inverter was built by connecting two TFT1 side-by-side with identical channel dimensions for both TFTs (*L*=0.2 mm, *W*=23.6 mm). In this inverter, the gate electrodes of both TFTs are common that serves as the input terminal V$_{in}$ as shown in Figure 10. Figures 11 a) and b) represents the inverter characteristic at supply drain voltage V$_D$ of + 1 V and -1V, respectively. When the supply voltage V$_D$ was kept +1 V, V$_{in}$ was varied from 0 to 2.0 V. Under this condition, TFT1 works as a p-channel transistor of a regular CMOS inverter while TFT2 operates as the n-channel device. Therefore, in positive gate bias, TFT1 operates in depletion mode and doesn't conduct current that results in a high V$_{out}$. As a result of this biasing, inverter works in the first quadrant and the output voltage V$_{out}$ vs. V$_{in}$ plot exhibit a maximum gain of 13 which is shown in Figure 11 c). On the other hand, if the $V_D$ is biased with -1 V and $V_{in}$ varies from 0 to -2.0 V (Figure 11 d), the inverter works with a gain of 12 which is represented in the third quadrant with n- and p-channel function exchanged between the two devices. The most advantageous side of this study is its low operating voltage with its reasonably good gain. As it was shown, this inverter needs only $|V_D| = 1.0$ V whereas, $|V_{in}|$ needs to vary from 0 to 2.0 V. Performance of such a solution-processed low operating inverter is rarely reported in literature.[51]

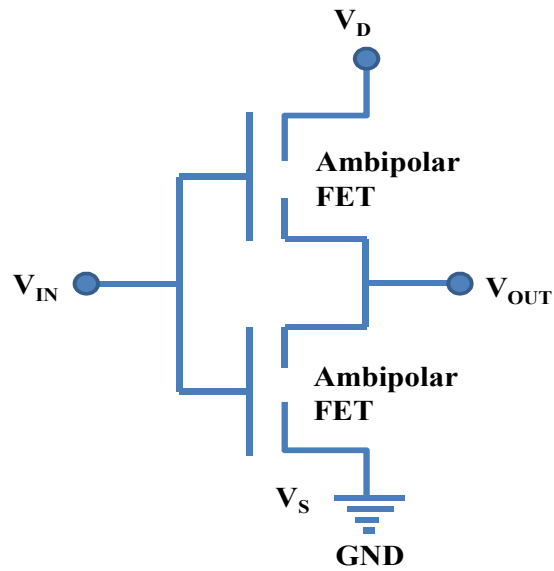

**Figure 10)** Schematic demonstration of the electrical networks for the inverter based on two identical ambipolar transistors

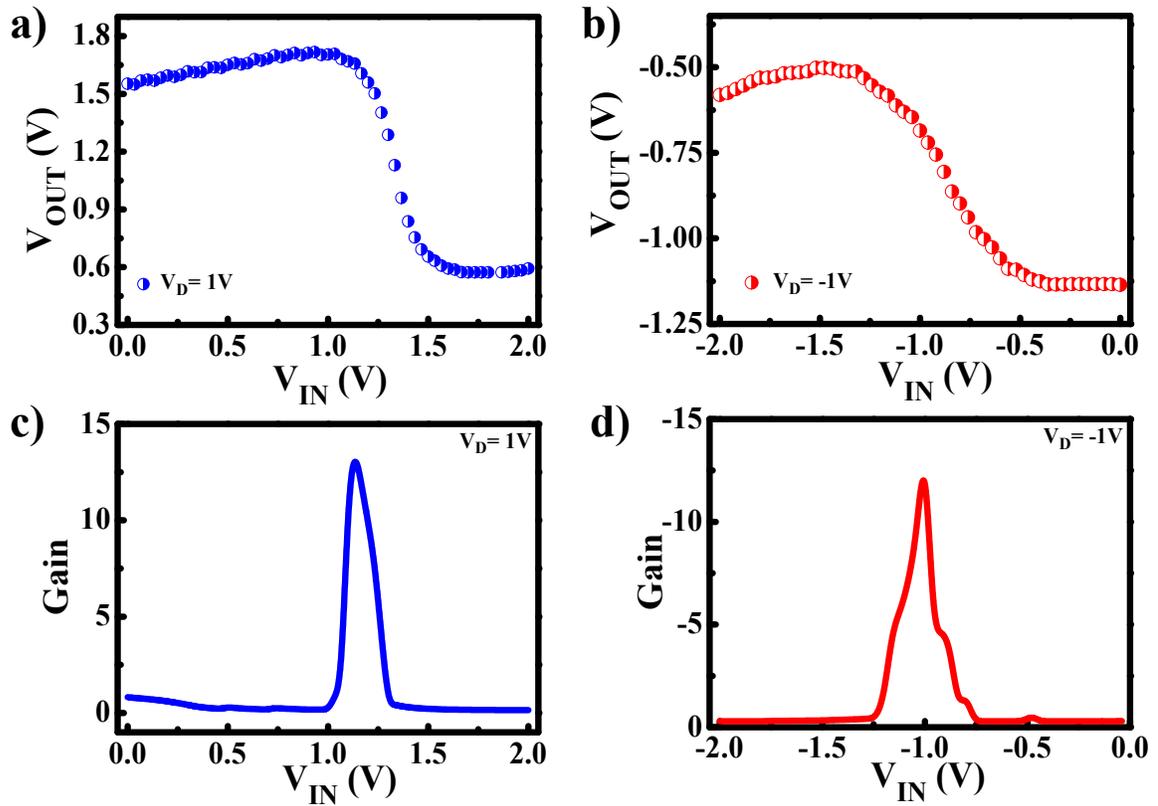

**Figure 11)** Inverter characteristics for **a)** first and **b)** third quadrants with supply voltages ($V_{DD}$) of ±1 V, respectively and corresponding gain of the complementary inverter **c)** first and **d)** third quadrants under the supply voltages ($V_{DD}$) of ±1 V.

## 4. Conclusion

In summary, $LiInO_2$ and $LiGaO_2$ ion-conducting oxide gate dielectrics those containt trivalent In and Ga respectively, have been developed by a low-cost solution-processed technique and successfully used as a gate dielectric for oxide-based ambipolar TFT. The optical property of $LiInO_2$ and $LiGaO_2$ thin films shows the films are highly transparent in the visible region with optical band gap of 3.6 and 5.5 eV respectively. The electrical conductivity of these dielectric thin films show the insulating nature which is a very essential condiction for using these ion-conducting oxide thin films as gate dielectric of TFT. During semiconductor ($SnO_2$) thin film fabrication on top of the ionic dielectric, those trivalent atoms allow p- doping to the interfacial $SnO_2$ layer to introduce the hole conduction in channel of TFT which has been identified with a reference TFT with $Li_2ZnO_2$ dielectric. Our comparative electrical data reveal that TFTs with $LiInO_2$ and $LiGaO_2$ dielectric are ambipolar in nature. However, TFT with $LiInO_2$ dielectric shows better performance with 1.0V operation voltage. This $LiInO_2$ dielectric based 1.0 V TFT shows balanced ambipolar TFT behavior with a high electron and hole mobility values of 7 $cm^2 V^{-1} s^{-1}$ and 8 $cm^2 V^{-1} s^{-1}$ respectively with an on/off ratio >$10^2$ for both operations which has been utilized for low-voltage CMOS inverter fabrication. Overall, this new technique of low operating volatage ambipoalr oxide TFT fabrication open up a new direction for developing high performance ambipolar TFT.

**Acknowledgements**

This work was supported by "Science and Engineering Research Board", India (EMR/2015/000689). Authors are grateful to Central Instrument Facility Centre, IIT(BHU) for providing the SEM and AFM measurement facility. Anand Sharma and Nitesh K Chourasia thanks IIT(BHU) for providing PhD fellowship. Nila Pal thanks "Science and Engineering Research Board" for providing Senior Research Fellowship. Sajal Biring acknowledges the financial support from Ministry of Science and Technology, Taiwan (MOST 105-2218-E-131-003, 106-2221-E-131-027, and 107-2221-E-131 -029 - MY2).